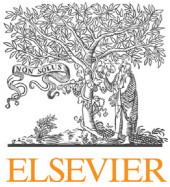
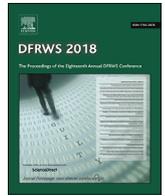

# Digital forensic investigation of two-way radio communication equipment and services

Arie Kouwen [a], Mark Scanlon [b, *], Kim-Kwang Raymond Choo [c], Nhien-An Le-Khac [b]

[a] *School of Computer Science, University College Dublin, Ireland*
[b] *Forensics and Security Research Group, University College Dublin, Ireland*
[c] *Department of Information Systems and Cyber Security, University of Texas at San Antonio, USA*

## ABSTRACT

Historically, radio-equipment has solely been used as a two-way analogue communication device. Today, the use of radio communication equipment is increasing by numerous organisations and businesses. The functionality of these traditionally short-range devices have expanded to include private call, address book, call-logs, text messages, lone worker, telemetry, data communication, and GPS. Many of these devices also integrate with smartphones, which delivers Push-To-Talk services that make it possible to setup connections between users using a two-way radio and a smartphone. In fact, these devices can be used to connect users only using smartphones. To date, there is little research on the digital traces in modern radio communication equipment. In fact, increasing the knowledge base about these radio communication devices and services can be valuable to law enforcement in a police investigation. In this paper, we investigate what kind of radio communication equipment and services law enforcement digital investigators can encounter at a crime scene or in an investigation. Subsequent to seizure of this radio communication equipment we explore the traces, which may have a forensic interest and how these traces can be acquired. Finally, we test our approach on sample radio communication equipment and services.

## 1. Introduction

Since Guglielmo Marconi (1874–1937) made a radio connection over a few kilometres in 1895, there have been many developments in the world of radio equipment. Over the past decade, a trend is noticeable in commercial radio-equipment increasingly switching from analogue to digital. When speaking of digital two-way radios, this digital equipment has several facilities which are commonly found on cellphones, such as address books, short message services, call logs, GPS, telemetry (the automatic measurement and wireless transmission of data from remote sources), etc. Today, telemetry applications include measuring and transmitting data from sensors located in vehicles, smart meters, power sources, robots, and even wildlife in what is commonly referred to as the Internet of Things.

Two-way radio is often referred as "professional mobile radio", "private mobile radio" (PMR), or "land mobile radio" (LMR), and colloquially referred to as walkie-talkies. Two-way radios most often use the Very High Frequency (VHF) and Ultra High Frequency (UHF) bands. Portable two-way radios have a communication distance of a few kilometres when directly transceiving to/from each other but when they make use of radio repeaters or Radio over IP (RoIP) the distance is almost unlimited. Overby and Cole outlines a comparison of telephony and two-way LMR services and a comparison of radio over IP and LMR IP trunking (Overby and Cole, 2008). Radio technologies are also used in Public Protection and Disaster Relief (PPDR) emergency response systems (Barbatsalou et al., 2014).

Push-To-Talk (PTT) services are increasingly used to communicate between different kinds of devices and radio equipment. There are numerous smartphone applications designed specifically for this purpose. One example of this is the WAVE Communicator application from Motorola, which uses their "WAVE Work Group Communications Solution". This makes it possible to directly communicate between two-way radios and smartphones. Some smartphones have a special PTT button, a soft-button on the screen, or reassigns an existing button, e.g., using the volume down to act as a PTT button. This makes radio communication equipment more popular among organisations that need group communication facilities that are independent of public communication infrastructures in case of an outage of this public infrastructure.

* Corresponding author.
*E-mail addresses:* arie.kouwen@ucdconnect.ie (A. Kouwen), mark.scanlon@ucd.ie (M. Scanlon), raymond.choo@fulbrightmail.org (K.-K. Raymond Choo), an.lekhac@ucd.ie (N.-A. Le-Khac).

The market for two-way radio is growing worldwide. According to Hytera (one manufacturer of Private Mobile Radios), there was a 100% market growth from 2014 to 2015.Land Mobile Radio (LMR) Systems (TETRA, Project 25, dPMR, DMR and TETRAPOL) market is expected to grow to $42 billion by 2022 (Acute Market Reports, 2016). Although communication possibilities such as cellphones, smartphones, phone lines, leased lines, and Internet exist, the infrastructure needed for these communication methods can experience blackouts. Certain parts of a nation's infrastructure are often considered critical as a failure or disruption can have serious consequences (Klaver et al., 2013). Because of this, mission-critical organisations resort to two-way radio, with which they can continue communication in case of infrastructural issues (Baldini et al., 2014).

Law enforcement agencies do not have much expertise with radio-equipment such as HF-, VHF- and UHF-Transceivers, Packet Radio, Digital Mobile Radio, Software Defined Radio (SDR), etc. Traditionally, this did not present a problem as radio equipment did not have much forensic value. However, modern systems use a variety of digital techniques such as digital speech- and data-channels, programming, and GPS. Data communication such as email, chat, location tracking or telemetry are also possible and have long showed promise for vehicular communication (Feher, 1991). Digital Mobile Radio can connect to backbone-equipment, which can further connect radio-transceivers with each other or, via internet-links, to other remote areas anywhere in the world. Furthermore, telecom operators and radio equipment rental companies in European countries are also offering PTT services.

To get a general insight into the existing knowledge of law enforcement digital experts, a questionnaire was sent out to Dutch Experts eXchange (DEX) members. The knowledge of the Dutch digital experts gives a mixed view. There are digital experts who have already encountered radio communication equipment and/or services at a crime scene and a number of cases where criminals used it to aid in the execution of their crimes, and those who have not. Of the 47 respondents, 12 had previously encountered radio communications in their cases; digital two-way radios were encountered in 7 cases, analogue two-way radios in 6 cases, smartphones with Push-To-Talk features in 4 cases, VHF/UHF transceivers in 3 cases, shortwave transceivers in 2 cases, WiFi two-way radios in 2 cases, data communication modem connected to a radio transceiver in 1 case, and Software Defined Radio in 1 case. 68.1% of the respondents "do not know" or "have little knowledge" of the intricacies of modern radio communication equipment. The majority of respondents (82.6%) identified that they would like to know more about the subject. Furthermore, the answers showed that radio communication is in use by criminals who obviously use it to hide their communication from being picked up by law enforcement. There were also cases in which radio communication was used in normal business situations. The overall results showed that more research for the subject was needed.

Because of the existence of this equipment and these services, it is likely that police will encounter this equipment in more and more cases, especially with the opportunity for criminals to leverage the technology in combination with other devices. Law enforcement are continuously battling to keep up with new technologies and devices (Lillis et al., 2016), while dealing with current digital evidence backlogs (Scanlon, 2016) However, there is very little research both in literature and by practitioners on digital forensic traces in radio communication devices. Therefore, in this paper, we present the forensic acquisition and analysis of radio communication equipment and services, and a workflow to aid investigation. We also evaluate the possibility of using popular forensic tools to acquire artefacts from radio communication equipment. We also test our approach with different scenarios and propose a workflow for radio device investigation.

### 1.1. Problem statement

Radio communication equipment is migrating from analogue devices to digital devices with new features commonly found in smartphones, such as call logs, address books, short messages, data communication and GPS. Because of these digital features and other benefits, radio communication equipment is increasingly used today. In addition, telemetry applications can make use of radio communication equipment and is in use by companies and organisations that need control data for their objectives. If digital investigators encounter digital radio communication equipment, it is necessary to have knowledge about the radio communication equipment, radio infrastructure, and associated services. However, there is little literature available on the topic.

When digital experts do not investigate digital radio communication equipment, valuable evidence may be neglected. This can be the case when a digital expert is not aware of the features of radio communication equipment and their networks. The following research questions are defined to get an insight into the current general knowledge level of digital investigators, the radio communication equipment and its users, where evidence can be found and how to get this evidence.

1. Who are the users of radio communication equipment?
2. Which equipment used for radio communication is worth to be investigated and which digital forensic traces may exist in radio communication equipment?
3. Is it possible with popular digital forensic tools to acquire radio communication equipment?
4. How can forensically interesting data in the radio communication equipment be acquired?
5. Where can other possible traces of evidence be found?
6. Forensic acquisition and analysis

## 2. Background

There are several manufacturers of digital radio communication equipment and software. The common brands are Motorola, Hytera, Sepura, Kenwood, ICOM, Vertex, Yeasu, Harris, Tyt Radio, amongst others. They offer portable and mobile two-way radios, repeaters and all kind of accessories such as headsets and remote speaker microphone (RSM) sets.

### 2.1. Features of digital radio equipment

Digital two-way radios both mobile and portable make use of one of the aforementioned standards. These digital standards make it possible, besides regular voice communication, to use many additional features and options. The features and options include:

- Radio-ID: This identifies the radio unit in the network. With TETRA it is called an Individual Tetra Subscriber Identity (ITSI) and consists of 3 individual numbers: Tetra Mobile Country Code (TMCC), Tetra Mobile Network Code (TMNC) and the Short Subscriber Identity (SSI). With DMR a Radio-ID and optionally a Radio Alias can be programmed.
- Talkgroups: Users/radios connected to the same talkgroup can communicate with each other. A user can switch to another

talkgroup and also a dispatcher can switch the radio remotely to another talkgroup.
- Zone: this allows users to organise channels conveniently. Each zone can support a certain number of channels. With the channel knob or via the menu the user can select a channel in the zone. If more than one zone is programmed into the radio, the user can change the zone via the programmable keys or menu (if the zone menu option is checked and zone programmable buttons is set).
- Private Call: normally a group call is established but users can also make a private call by entering a name or radio-ID on their radio and than PTT.
- Trunk or DMO operation: a radio can be used in Trunked Mode Operation (TMO) or in Direct Mode Operation (DMO). In TMO, a network controller (computer) assigns the correct channel and other parameters to the radio. Radios can communicate directly with each other in DMO.
- Roaming: is designed to have the digital radio automatically select the best channel if Receive Signal Strength Indicator (RSSI) of the current channel falls below a defined level as the radio moves throughout the coverage area of a group of repeaters that carry the same Talk Groups on the same time slots.
- Encryption: all kind of encryption methods exists, from entering a software encryption key to installing an extra printed circuit board called Option Board in the radio.
- Address Book: the address book contains radio-ID numbers with their corresponding user or department names.
- Status Messages: a digital radio can have pre-programmed short status messages (1–16 characters). For instance, the radio user can use these messages as a reply on a work order.
- Short Message Services: short messages can be send and received similar to cell phone SMS messages.
- Radio Check: with radio check you can check whether another radio is on or off.
- Emergency call: when an emergency call is initiated, it is clearly visible and audible at the receiving end. Whether it is coming into a dispatcher workstation or another radio, team members know immediately that there is an emergency. These calls get a higher priority than regular calls.
- Lone Worker: is designed for persons who work alone. If the user cannot operate the radio within the pre-set time due to emergency, the radio will make an emergency alarm automatically to inform user's colleagues or the control centre for help.
- Remote Monitoring: with remote monitoring a radio can be placed in listening mode, it then will send all sound it picks up with its microphone.
- Password: a password can be configured. After switching on the radio a password must be entered in order to operate the radio if it is password enabled.
- Remote enable/disable: if remote enable/disable is configured, the radio can remotely be enabled or disabled. Disabling may also result in wiping the data in the radio.
- Call Alert: makes it possible to send an audible alarm to a remote radio.
- Rental: a rental period can be programmed in the radio. When the rental period is over, the radio cannot be used any further.
- Telemetry: by configuring parameters of this function, users can control or enquire status of the target radio's General-Purpose Input/Output (GPIO) port, and can send status of the user's radio GPIO port. With this all kinds of processes can be controlled.
- GPS: with the GPS option the radio can send it's location to a server. This can be done automatically or manually by the user.
- Keyboard Lock: with keyboard lock active only the pre-programmed functions of the radio are active and can be changed without entering the keyboard lock deactivation key sequence. Often only the volume setting can be changed when keyboard lock is active. With keyboard lock active radio settings cannot change without interaction of the user.
- Covert Mode: this feature makes the radio totally silent. In this mode, any user interface indication on the radio is prohibited.

## 2.2. Accessories for radio communication equipment

Mobile and portable two-way radios can use a variety of accessories such as remote speaker microphones, remote speaker microphone with a build-in camera, GPS and voice recorder, neck loops in combination with an earpiece, Bluetooth wireless audio accessories, Bluetooth PTT-buttons, program cables and CPS. The speaker microphone may contain a micro-SD card, which can save pictures, videos and recorded audio.

## 2.3. Software Defined Radio

The first generation of SDR only explicitly needed radio components to exist in the hardware box and the rest of the communication was done with software. Nowadays, the hardware box not only has radio components on board but also computing capabilities. The latest generation of SDR acts like a server and the console to control the radio and can be hardwired or connected via a network. FlexRadio[1] is a manufacturer of both hardware and software to communicate with SDR. Small SDR receivers are also in the market. For example, RTL-SDR[2] is a very cheap device that uses a DVB-T TV tuner dongle based on the RTL2832U chipset. With this device, it is possible to receive and decode a variety of radio signals, as outlined on their website. Digital Speech Decoder (DSD)[3] is an open source software package that decodes several digital speech formats. It uses the `mbelib` library to synthesise the decoded digital speech. It does not however allow decoding of encrypted communications.

## 2.4. Push-to-talk applications

PTT has been used by cellular providers for many years. PTT allowed providers to fill their unused airspace by providing almost real-time communication (Chen et al., 2007). Since smartphones became popular in the market, several PTT applications have been developed. There are many applications that can communicate with each other acting like a two-way radio. Some of them also have the ability to connect to radio communication networks. An example of this is Motorola's WAVE solution. With the WAVE Mobile Communicator application installed, it turns a smartphone into a multi-channel PTT handset for fully secure real-time voice and text communications. Users can connect to other smartphones users or connect to radio users in a radio network. The Cisco Instant Connect solution[4] is another total communication solution with similar features. HAM radio amateurs often use a PTT application called Echolink. Other popular PTT applications are Zello, iTeamTalk, Vodafone PTT, KPN PTT, and GoPTT. There are also PTT applications that communicate via WiFi and/or Bluetooth. This makes it possible to create a private and secure communication environment when using the Wireless LAN's or Bluetooth's security features. An example of such an application is WalkieTalkie from Porchlight. A new product that will be entering the market soon is the DXBm[5] made by Fantom Dynamics. The DXBm is modular system, which allows the user to transform their smartphone or tablet into their own unique off-the-grid peer-to-peer network. With the DXBm and for instance a connected DMR module, a smartphone can also be a DMR two-way radio. The DXBm has other different modules available to fit every user's need including: UHF module 1 Watt and 2 Watt, Dual band module (VHF/UHF), 800Mhz module, DMR module, P25 module, Radio over IP (RoIP) module, and MDC-1200 module with encryption.

---

[1] https://www.flexradio.com.
[2] http://www.rtl-sdr.com.
[3] http://wiki.radioreference.com/index.php/Digital_Speech_Decoder_(software_package).
[4] https://www.cisco.com/c/en/us/products/collateral/physical-security/ipics-server-software/data_sheet_C78-728836.html.
[5] https://www.fantom.io/Products/DXBm/.

### 2.5. Users of radio communication equipment

In general, two-way radio is often used when it is required to have a fast direct group or private call communication with no call setup delays without making use of public infrastructures, such as GSM, UMTS etc. There are a wide variety of users of radio communication equipment. Radio communication equipment has users in public entities, such as public works, municipalities, embassies and ministries, security, medical services, and in private enterprises & businesses such as oil & gas, restaurants, resorts and hotels, events and concerts, radio rental companies, and ham radio amateurs.

Ham radio amateurs are people who have a license to experiment and transceive on several radio frequencies designated to amateur radio services (Kuznetsov and Paulos, 2010). For some frequencies, Ham radio amateurs have a primary status, which means that the frequency is primarily for Ham radio. For other frequencies, they may have a secondary status, which means that other users have primary rights. A special Ham Radio service is the Amateur Radio Emergency Service (ARES).

### 2.6. Which equipment used for radio communication is worth to be investigated and which digital forensic traces may exist in radio communication equipment?

Radio communication equipment can contain useful digital forensic traces. The aforementioned HF transceivers may have storage media installed that can contain settings and recorded voice. Also connected devices and computer applications can have stored data.

Radio repeaters do not tend to store much data. They typically function as a receiver on one frequency and transmit the received payload on another frequency. However it can be valuable to investigate the configuration, in particular the receive and transmit frequencies and if connected to a network, determination of the network configuration in use.

Portable and mobile two-way radio can contain configuration data that show the active settings of the radio. These settings can be downloaded from the radio and saved as a "code plug". This code plug can then be analysed by loading it into the Customer Program Software (CPS) application, software provided by the hardware vendor for configuring, reading, and writing data to/from the digital radio equipment, e.g., settings such as frequencies, talkgroups, contacts, power output, keylock, etc. The radio may have additionally stored call logs and message logs that may be valuable for an investigation. The two-way radio may be equipped with a camera and storage media, which can contain data to be investigated. Often accessories such as a remote speaker/microphone or Bluetooth devices are connected to portable two-way radios for PTT applications. The simpler versions of these remote speaker/microphone devices does not contain valuable digital traces. The more advanced versions can contain a camera, voice recorder, and storage media. Remote PTT buttons are also available in the marketplace. These are small buttons that connect via Bluetooth to a smartphone or a two-way radio. These devices do not contain much digital forensic evidence but when these devices are encountered at a crime scene it may be an indication that a two-way radio or a PTT service is involved.

### 2.7. Using popular digital forensic tools to acquire radio communication equipment

Popular mobile-focused digital forensic tools on the market are Cellebrite's UFED Physical Analyzer, Magnet Acquire, MSAB's XRY and Blackbagtech's Blacklight. These are used to acquire GSMs (Duvinage, 2009), smartphones, and navigation equipment. However, we could not find support for two-way radio devices from any of the above-mentioned applications. Requests for information were sent to Cellebrite, MSAB and Magnet Forensics and all three companies replied that they do not have any experience with two-way radio and do not plan to invest in it in the future.

CPS can do a read of a two-way radio when the correct, often vendor-specific, cables are used. The more recent two-way radios from various brands have a micro-USB connection that makes it possible to use a standard cable for doing a read of the radio. With the CPS software installed in a clean virtual machine, a read of the radio can be conducted, which provides all the configuration settings such as frequencies, network settings, and contacts from the radio. The use of a clean virtual machine is recommended to ensure the highest level possible of a forensically-sound workflow as the vendor software is not designed to be forensically-sound.

In the daily work of a digital investigator, it is not always possible to do an extraction of every mobile device encountered. Sometimes, a mobile device is not supported by forensic software or there may be a risk of losing data when performing an extraction. In these cases, a manual extraction with the help of photography based device investigation software, e.g., Fernico's ZRT3.[6] This is the case with most two-way radios. With ZRT3, digital photos are taken from the screen contents of the mobile device. If needed the optical character recognition (OCR) capabilities of ZRT3 can be used to convert what is displayed on the screen to text. ZRT3 generates a digital or paper based report with the corresponding hash values of the digital pictures documenting the investigative process.

## 3. Acquisition of forensically interesting data from radio communication equipment

In this section, we describe how to get data from radio communication equipment through three experiments. The results are outlined in separate sub-sections below.

### 3.1. Winlink radio email

RMS Express is an email application developed by Winlink and was the focus of this investigative scenario. To guide investigators, we propose a workflow that could be followed in case of encountering a radio transceiver connected to a modem, as can be seen in Fig. 1.

A VHF radio was connected to a modem and subsequently connected to a computer. In this case, the transceiver was a dual-band VHF/UHF transceiver, but for radio email, any radio transceiver can be used. Using the steps in the flowchart: pictures were made of the equipment, cables and antenna. It was visible that the transceiver was receiving on the frequency 144.850 MHz. The transceiver was connected to an antenna that was placed indoors below the roof of a house. The equipment details used were as follows:

---
[6] http://www.fernico.com/ZRT3.aspx.

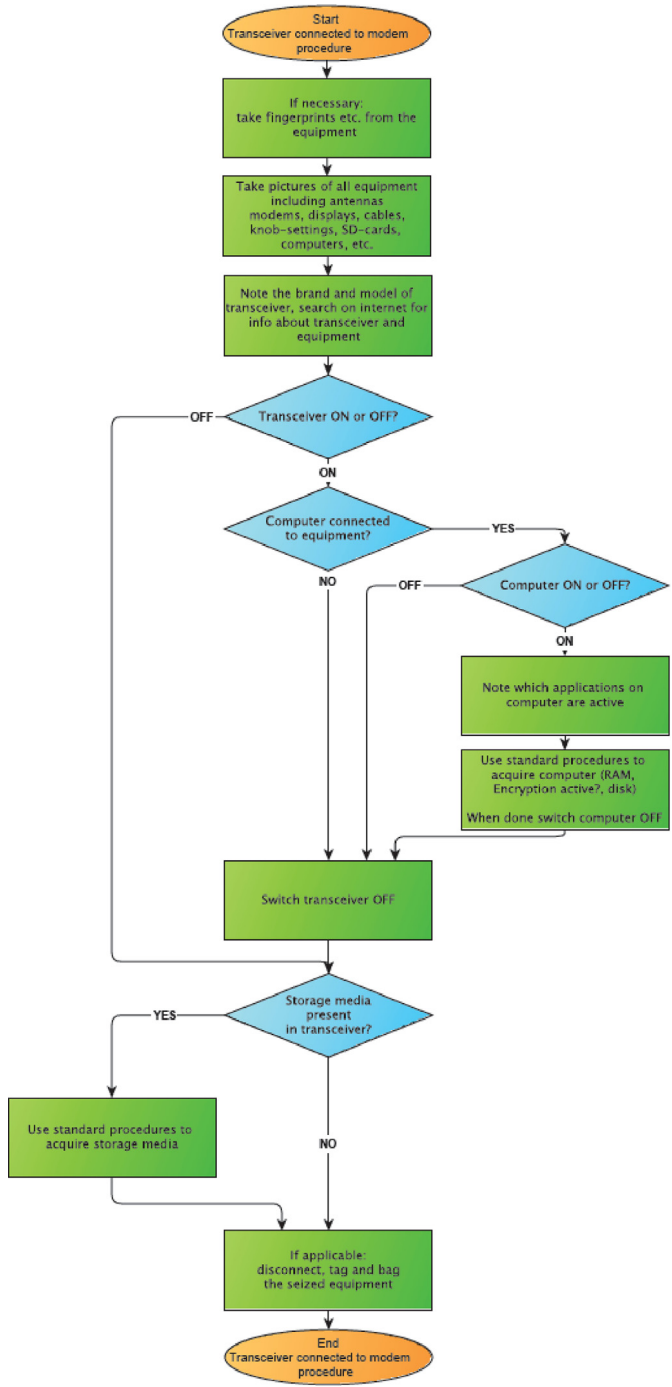

**Fig. 1.** Investigative workflow in the case of encountering a radio transceiver connected to a modem.

- A Kenwood TH-F7E VHF/UHF portable transceiver.
- A Diamond X-30 In-house antenna.
- A Tigertronics Signalink-USB Soundlink modem.
- A Windows 8 PC with RMS Express installed.

The transceiver had no storage media. The speaker/microphone port of the transceiver was connected to a Signalink soundmodem and the modem was connected to the PC via USB. On the computer display, two visible applications were active: a sound modem application and the RMS-Express email application with visible email-folders. Next, an external USB stick with FTK-Imager portable was inserted into the computer and a FTK ad1-image with MD5 and SHA1 checksum creation was made of the folders. Subsequently, the contents of the forensic image was analysed. In the folder "Logs", several log files were present. These log files contained date/time stamps with info about connection setup to a Radio Message Server, Central Message Server and message IDs that are sent or received, as can be seen in Fig. 2.

Sent and received messages with their corresponding timestamps in UTC40 were recoverable. The content of the messages are stored in the folder "Messages" and their filenames correspond with the message IDs and have a MIME type (Multi-purpose Internet Mail Extension). MIME is a protocol used to transport non-text information across the Internet.

### 3.2. Hytera PD-785G portable two-way radio

The Hytera PD785G is a portable radio with both analogue and digital radio. The digital radio uses the DMR protocol. The procedural steps that could be performed in case of encountering a two-way radio can be seen in Fig. 3. The antenna of the radio was detachable and was removed to prevent remote disabling. It was visible that the transceiver was active on the channel RPTR-1_TG9.2 in zone RePeaTeR-1. The channel-knob was set on position 1. As mentioned in the device's online manual, the correct programming cable (Model PC38) was purchased. A clean virtual Windows machine with the Customer Programming Software (CPS) version V7.06.02.006 was used. While the radio was connected, a read of the radio was performed with CPS and the resulting code plug file, named `Hytera_PD785G.rcdx`, was made read-only and saved. Both the MD5 and SHA1 hashes were calculated. Not all data could be read with the CPS software, i.e., non-configuration settings data such as call logs, short messages, etc. Therefore, a manual read of the missing items was done with the help of Fernico ZRT3. After finishing the manual read, the resulting ZRT3 files were acquired with FTK-Imager and from that resulting image, the corresponding hashes were calculated. The final step was to power down the radio and remove the battery. A picture was made of the inside showing brand, model, serial number, and frequency range.

To analyse the results the code plug file, `Hytera_PD785G.rcdx` was loaded into the CPS software. The identifying items of the radio: serial number, model, frequency range, radio alias and Radio ID were recoverable, which may be valuable for an investigation. To analyse the results of the manual read out, the Fernico ZRT3 report was examined. The anonymised Radio-ID and Radio-Alias were visible.

### 3.3. Motorola WAVE smartphone Push-To-Talk application

The WAVE Communicator PTT application was installed on an iPhone to be investigated. An extraction with Cellebrite UFED was conducted and the iPhone was also investigated with Blacklight from Blackbagtech. Data that could not be retrieved with UFED and Blacklight was read out manually with the use of ZRT3 from Fernico.

In this case, a demo account of Motorola's WAVE solution was used. Because of the risk of loosing data when doing an acquisition with forensic software, a manual acquisition was first conducted with Fernico's ZRT3. Subsequently, FTK-Imager was used to make a logical image of the ZRT3 files. Also, hash values of the logical image were created and verified. The ZRT3 report displayed the logged on WAVE User-ID, GPS location information, call logs, and text messages.

The active channel/talkgroup `TalkGroup-A` is shown in Fig. 4 (a) and the list of online users of this talkgroup PTT 1 and PTT 3

```
File List
Name                                    Size    Type              Date Modified
$I30                                       4    NTFS Index All... 29-3-2016 18:42:36
RMS Express 20160328.log                  24    Regular File      28-3-2016 19:54:00
RMS Express 20160329.log                  13    Regular File      29-3-2016 19:13:59
RMS Express Autoupdate 20160328.log       10    Regular File      28-3-2016 19:45:26
RMS Express Autoupdate 20160329.log        5    Regular File      29-3-2016 19:04:18

2016/03/29 19:04:06 1.3.10.0 -----  RMS Express version 1.3.10.0 starting  -----
2016/03/29 19:04:11 1.3.10.0 PAC *** Starting WL2K packet session...
2016/03/29 19:04:11 1.3.10.0 PAC *** Initializing TNC-X; port COM73; 9600 baud
2016/03/29 19:04:11 1.3.10.0 PAC *** Initialization complete
2016/03/29 19:04:11 1.3.10.0 PAC *** Ready
2016/03/29 19:04:14 1.3.10.0 PAC *** Starting to call PI8NHN-10
2016/03/29 19:04:14 1.3.10.0
2016/03/29 19:04:14 1.3.10.0 PAC *** Opening serial port COM73; 9600 baud; TNC-X
2016/03/29 19:04:14 1.3.10.0 PAC *** Connecting to PI8NHN-10
2016/03/29 19:04:17 1.3.10.0 PAC *** Connected to PI8NHN-10 at 2016-03-29 19:04:17
2016/03/29 19:04:18 1.3.10.0 PAC Welkom, RMS PI8NHN-10 Relay-Normal , PL:256, MF:4, MS:3
2016/03/29 19:04:34 1.3.10.0 PAC [WL2K-3.2-B2FWIHJM$]
2016/03/29 19:04:35 1.3.10.0 PAC ;PQ: 71386046
2016/03/29 19:04:35 1.3.10.0 PAC SanDiego CMS via PI8NHN >
2016/03/29 19:04:35 1.3.10.0 PAC       ;FW:
2016/03/29 19:04:35 1.3.10.0 PAC       [RMS Express-1.3.10.0-B2FHM$]
2016/03/29 19:04:35 1.3.10.0 PAC       ;PR: 95555287
2016/03/29 19:04:35 1.3.10.0 PAC       ; PI8NHN-10 DE        (JO22JQ)
2016/03/29 19:04:35 1.3.10.0 PAC       FC EM 2MK510YUDQLG 233 202 0
2016/03/29 19:04:35 1.3.10.0 PAC       F> A6
2016/03/29 19:04:53 1.3.10.0 PAC FS Y
2016/03/29 19:04:53 1.3.10.0 PAC *** Sending 2MK510YUDQLG.
2016/03/29 19:04:58 1.3.10.0 PAC FC EM PYXZC53QVQYJ 736 477 0
2016/03/29 19:04:58 1.3.10.0 PAC *** Completed send of message 2MK510YUDQLG
2016/03/29 19:04:58 1.3.10.0 PAC *** Sent 1 message.  Bytes: 230,  Time: 00:04,  bytes/minute: 2776
2016/03/29 19:04:59 1.3.10.0 PAC F> 25
2016/03/29 19:04:59 1.3.10.0 PAC     FS Y
2016/03/29 19:05:04 1.3.10.0 PAC *** Receiving PYXZC53QVQYJ
2016/03/29 19:05:08 1.3.10.0 PAC *** PYXZC53QVQYJ - 751/487 bytes received
2016/03/29 19:05:08 1.3.10.0 PAC *** Bytes: 523,  Time: 00:03,  bytes/minute: 9677
2016/03/29 19:05:08 1.3.10.0 PAC     FF
2016/03/29 19:05:11 1.3.10.0 PAC FQ
2016/03/29 19:05:11 1.3.10.0 PAC *** --- End of session at 2016-03-29 19:05:11 ---
2016/03/29 19:05:11 1.3.10.0 PAC *** Messages sent: 1.  Total bytes sent: 230,  Time: 00:54,  bytes/minute: 253
2016/03/29 19:05:11 1.3.10.0 PAC *** Messages Received: 1.  Total bytes received: 523,  Total session time: 00:54,  bytes/minute: 575
2016/03/29 19:05:28 1.3.10.0 PAC *** Disconnected at 2016-03-29 19:05:28
2016/03/29 19:05:28 1.3.10.0 PAC *** Disconnect reported.
```

**Fig. 2.** Content of RMS express log file.

is displayed in Fig. 4(b). The WAVE Communicator can also use GPS to show the locations of the users. Each user can enable or disable this on the smartphone application. Fig. 4(c) shows the location of user PTT 3. The WAVE Communicator application also records the voice calls history. Fig. 4(d) shows some of the calls that were made by user PTT 1 and PTT 3. This screenshot also shows that PTT 1 and PTT 3 had a private call 40 s ago. The WAVE system also has the ability to send and receive text messages. Fig. 4(e) shows some of the messages user PTT 1 has sent and the message user PTT 3 has sent. These logs only remain visible as long as the application is active. Whenever the application is closed, all of the aforementioned data is gone.

After the manual acquisition, the iPhone was acquired with Cellebrite's UFED; both file system and logical dumps were taken. Next, the results were analysed with Cellebrite's Physical Analyzer. Searches with the keywords wave and twistedpair (the application's developer) resulted in several hits, as can be seen in Fig. 5. Analysing the hits from the keyword search for 'wave', one notification and five files were visible. Next, the iPhone was connected to the Blacklight application from BlackBagtech and the same searches were performed. Blacklight also gave hits on the keywords. Both Physical Analyzer and Blacklight pointed to a filename `com.twistedpair.wavetc.plist` in the folder `/mobile/Applications/com.twistedpair.wavetc/Library/Preferences/`. The plist file was opened with Apple's Xcode, which showed the cached username, password and server for the application. No traces of call history or text messages were found. Because of this, it is necessary to do a manual acquisition to ensure comprehensive data recovery.

### 3.4. Other possible traces of evidence

When radio communication devices are used in an investigation, it might be worthwhile to search for other traces of evidence. These additional sources are highlighted in this section for completeness, but their investigation is beyond the scope of the paper. Often the providing agency may keep records of the users of a frequency. This means that when the transceiver frequencies of a radio communication device are known, the Radio Communication Agency can tell who the possible users are. They may also have information about the radio signals they monitor in real time or have saved in the history database.

In a mobile radio network, there may also be dispatch equipment installed. Dispatch equipment can have log files with call logs, text messages, GPS data, or recorded voice. The recorded voice may also exist on a separate voice recorder server. These log files and recorded voice can be valuable for the investigation. It is important to know that in particular the recorded voice may have a short retention time. This means that it is necessary to act as soon as possible to get the recorded voice and other relevant data in time. For example, the mobile radio network company, Entropia, uses a retention time of one week. The radio email system, Winlink, may have log files on the Central Message Servers (CMS). According to the system administrator, there are only log files of connections that are made via Telnet. If needed these log files can be acquired via a MLAT. The Radio Message Server keeps several log files of the communications it provides via radio and Internet. However, there is a limited time of 2 weeks that these log files exist on the RMS.

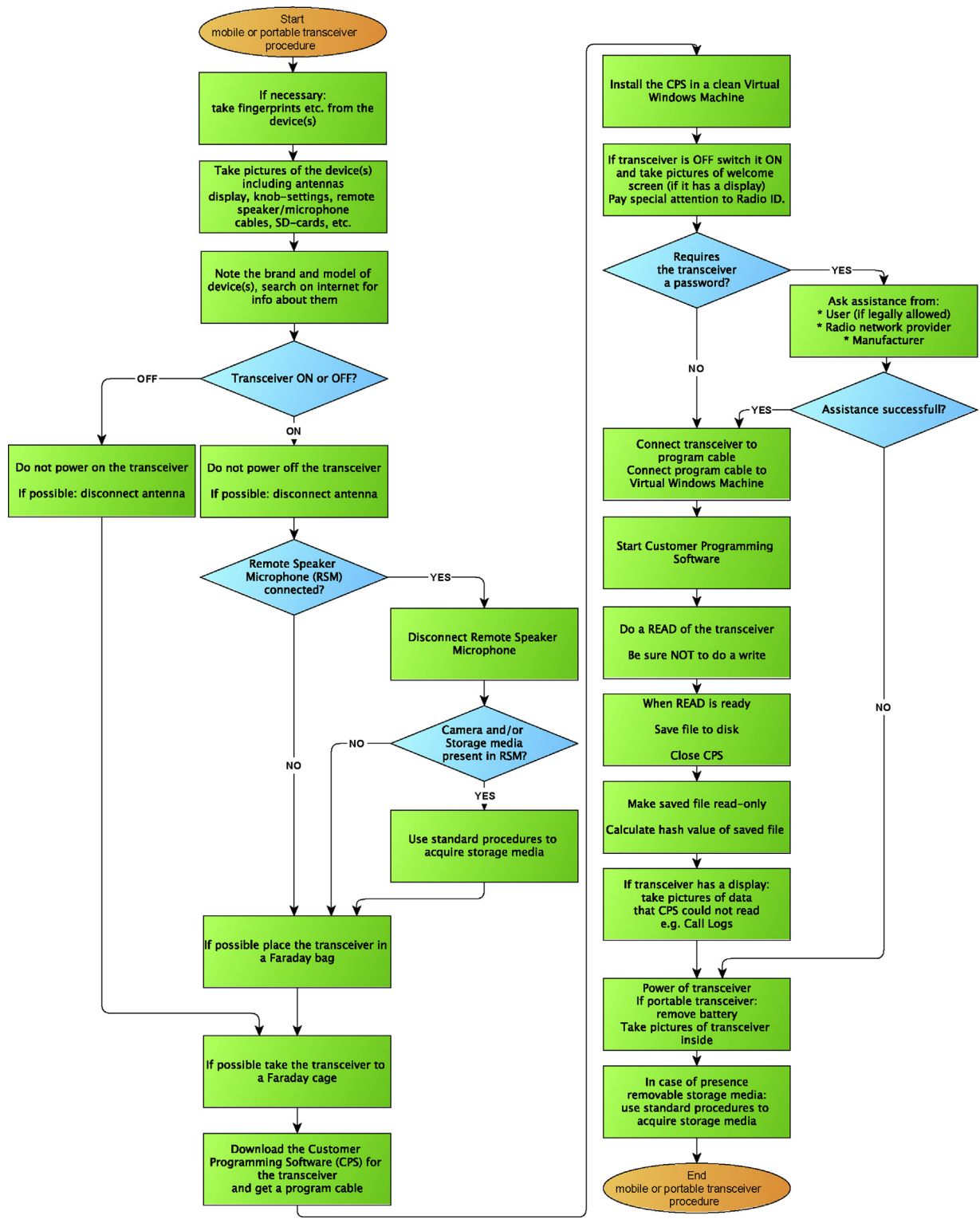

**Fig. 3.** Investigative workflow for processing mobile or portable two-way transceiver.

A PC may have an installed PTT application or may use a web browser to use PTT facilities. Examples of these are the Motorola WAVE Desktop Communicator and WAVE Web Communicator. These applications keep logs which may also prove valuable to the investigation.

As mentioned previously, digital radio devices make use of a Radio-ID for communicating with each other. Ham-radio amateurs have connected their repeaters worldwide to Ham DMR networks. A popular DMR network is called the Brandmeister network[7]. This network offers a dashboard portraying useful information about repeaters. It also has a last heard section that shows the last established connections in the DMR radio network.

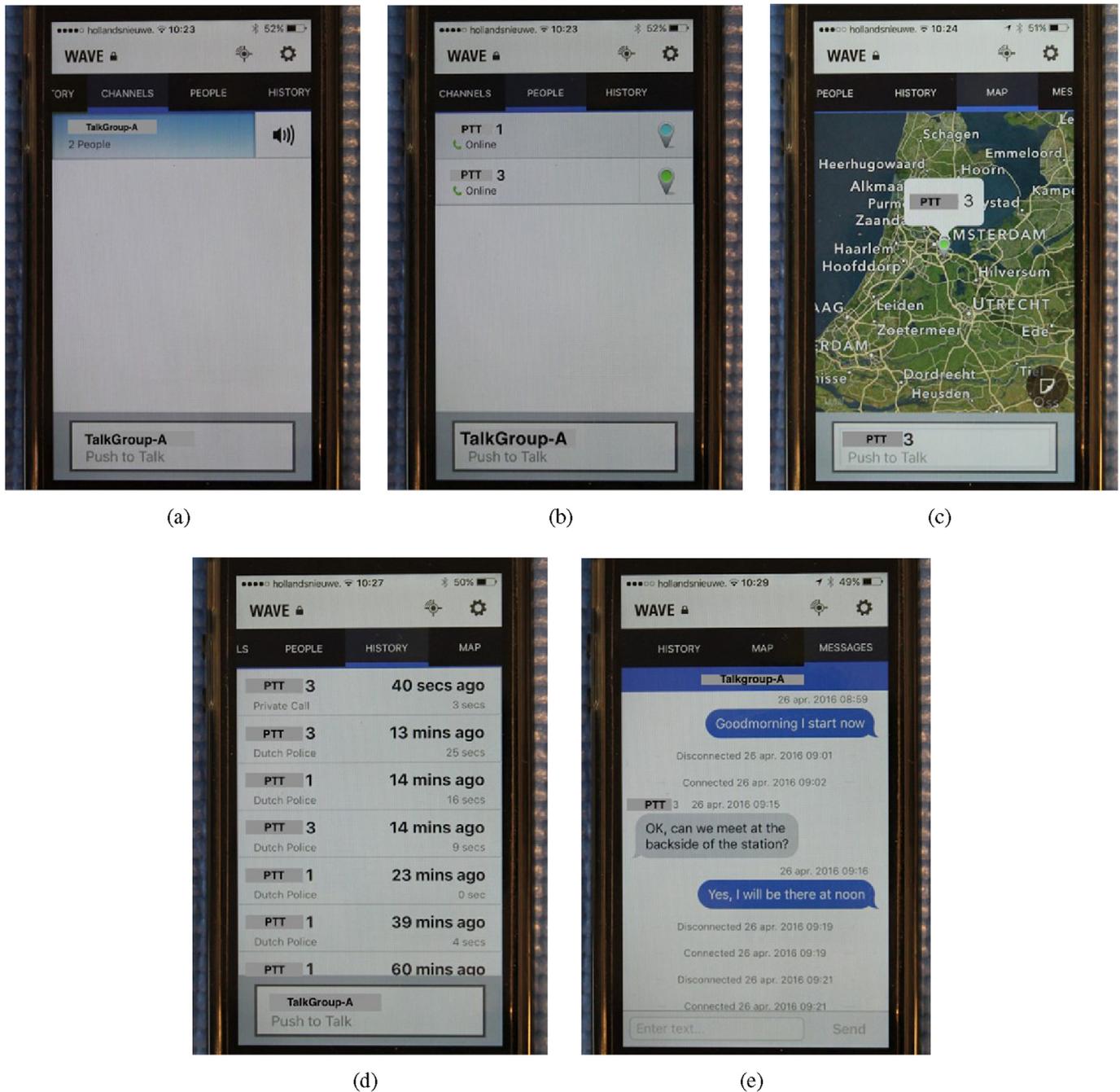

**Fig. 4.** Screenshots of the Motorola WAVE iPhone Application.

### 4. Analysis and discussion

The motivation for this research was the lack of knowledge regarding radio communication equipment and services and thus the risk of investigators missing out in an important source of evidence. Knowledge of this equipment, the radio infrastructures and services is important to find possible evidence.

The market for Land Mobile Radio and PTT services is still growing and many enterprises and organisations are using two-way radio. The radio communication equipment used varies from hand held devices to special radio equipment used for connecting computers. Because of the existence of many radio communication devices, it was not possible to research all products for the purpose of this paper. Radio communication devices can be recognised because every radio device needs an antenna which is most often outside the device connected directly or via a coax cable. Because of the large amount of users of radio communication equipment it makes sense to expect this radio equipment at certain crime scenes. It is therefore good practise to pay attention to the possible

---

[7] https://bm.pd0zry.nl/index.php/Main_Page.

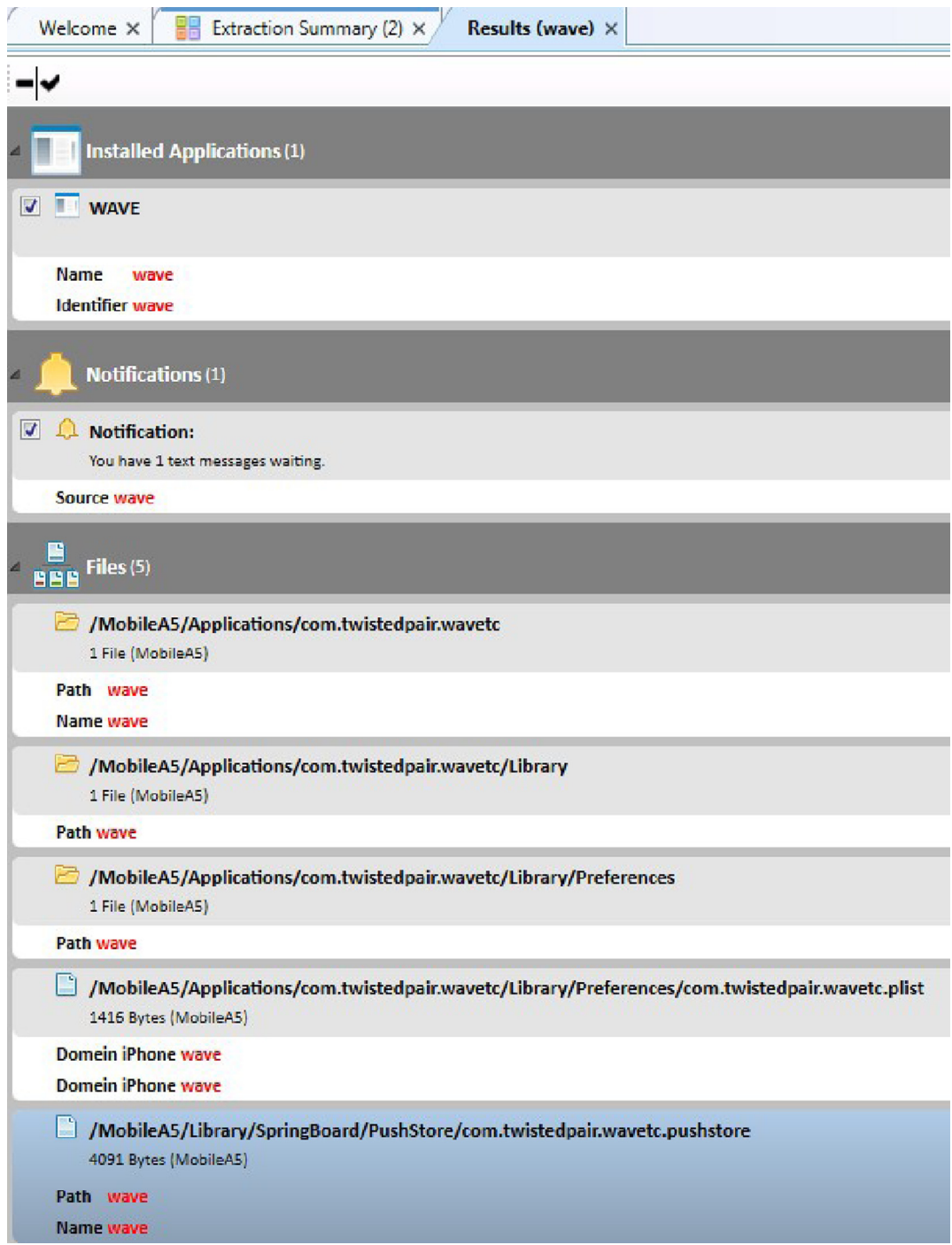

**Fig. 5.** Physical Analyzer Search Results for the Keyword 'Wave'. *MobileA5* refers to the name of the acquired iPhone.

presence of radio communication equipment and to take adequate action when they are found.

Websites like Alibaba.com and Dealextreme.com sell radio communication equipment from DMR radios to HF radios. These radios are most often used in the Ham Radio frequency bands where normally a license is required. However, the above mentioned websites do not strictly limit themselves to selling to Ham radio amateurs only. Furthermore the worldwide interconnected DMR radio repeaters used for Ham radio can be used for connections between criminals, eliminating the need for utilisation of cellphones due to their relative ease of investigation (Hoy, 2014). For getting a Ham radio DMR-ID you need to register and send some prove of owning a Ham license, but the DMR network does not check who you really are when you make a voice or data call. No authentication is taking place so therefore, it is possible to use a random DMR radio-ID and make connections.

The monitoring of radio frequencies in the Netherlands is being done by the Radiocommunications Agency Netherlands (Dutch:

Agentschap Telecom). However, they only monitor the use of the frequencies and not the content of the radio communication. Neither does the Dutch Police. As a result, it is conceivable that radio communication equipment may be used for communication between criminals.

There are several affordable SDR peripherals on the market ranging from USB TV tuners, capable of receiving radio signals, to bespoke hacker-focused SDR hardware, e.g., HackRF One,[8] USRP,[9] etc., capable of receiving and transmitting radio signals. These two-way peripherals requires the special attention of law enforcement digital experts. With two-way SDR hardware, the device can act as a complete radio communication transceiver. It also facilitates signals to be recorded and replayed (Anderson et al., 2015). For instance, when the transmission of a remote car key has been recorded, it is then possible to open or close doors of the car by doing a replay of the earlier recorded radio transmission (Heinäaro, 2015). Fortunately, some of the car manufacturers are aware of these possibilities and are starting to take precautions against this.

In cases of radio communication equipment being connected to computers, it is not difficult to get relevant data such as log files or emails from the computer using regular digital forensic techniques. Important log files from servers, such as dispatcher servers, voice-recording servers or GPS location servers, can mostly be acquired from the servers involved.

## 5. Conclusion and future work

Information for this paper has been searched for and found regarding forensic digital traces in radio communication equipment and services. As the time of writing, the literature survey yielded information about the digital radio techniques, standards, products and users, but did not give any information about how to acquire data from digital two-way radios. Currently there is no way of acquiring the two-way radio devices with the standard mobile forensic tools like Cellebrite's UFED, MSAB's XRY and tools from Magnet Forensics. For getting the data from transceivers and connected computers and getting data from a two-way radio, two procedural flowcharts have been provided, which may help as a guide to get the data from radio devices and connected computers. With these procedures, it was possible to successfully get the data from the radio communication equipment. Much of the work presented requires manual interaction with the devices and significant room is left for improving the efficiency and automation of the process (In de Braekt et al., 2016).

When a digital investigator has proper knowledge of radio communication equipment and services, the digital investigator can then search for relevant forensic evidence in both the equipment or servers in the radio communication infrastructure. The evidence can sometimes be extracted from the devices and sometimes it is necessary to go to a radio network service provider to acquire the evidence from connected servers. It is important to check the radio communication equipment for digital traces and look for settings, used frequencies, connected accessories with possible storage media, connections to other users, servers, gateways, dispatch applications, etc. Due to the Internet Protocol-capable (IP) feature of digital two-way radio equipment, one might reasonably expect more data communication applications to be available on the market. Almost all manufacturers of digital two-way radio equipment offer IP-based facilities and applications, but users need their proprietary cables and applications.

The conclusion is that law enforcement digital experts must investigate digital radio equipment. There are ways of getting the data from this equipment, however it depends on the kind of equipment if and how this can be done. Industry is moving towards integrated devices, so it is to be expected that more often radio communication equipment combined with smartphones will be encountered at crime scenes. Therefore it is important to get access to the radio devices with forensic tools. That is why API possibilities, JTAG and chip-offs have to be researched and developed, in a similar manner to mobile phone forensics (Alghafli et al., 2012).

It may be possible that criminals will find a way of using radio communication equipment, such as bespoke hacker SDR hardware and GNU radio, to form a threat for radio communication networks including the upcoming Long Range (LoRa) networks. Further research in this equipment and software is required to get a better insight into the features and possible threats and training should be developed to support law enforcement encountering this equipment in the field (Hitchcock et al., 2016).

---

[8] https://greatscottgadgets.com/hackrf/.
[9] https://www.ettus.com/.